\documentclass[12pt]{article}
\usepackage{graphicx}
\usepackage{enumerate}
\usepackage{amsmath}
\usepackage{amsfonts}
\usepackage{amssymb}
\usepackage{amsthm}
\usepackage[dvipsnames]{xcolor}
\usepackage[
  bookmarks=false,
  pdfpagelabels=false,
  hyperfootnotes=false,
  hyperindex=false,
  pageanchor=false,
  colorlinks,
]{hyperref}

\newtheoremstyle{dotless}{}{}{}{}{\bfseries}{}{ }{}
\theoremstyle{dotless}

\newtheorem{rem}{Remark}
\newtheorem{defn}{Definition}
\newtheorem{lem}{Lemma}
\newtheorem{prop}{Proposition}

\newtheorem{ob}{Observation}
\newtheorem{ex}{Example}

\newcommand{\bt}{\begin{theorem}}

\newcommand{\et}{\end{theorem}}
\newcommand {\be}{\begin{equation}}
\newcommand {\ee}{\end{equation}}

\newcommand{\hsp}{\hskip 2em}

\newcommand{\ol}{\overline}
\newcommand{\noi}{\noindent}

\def \qed {\hfill \vrule height6pt width6pt depth0pt}
\def \noi{\noindent}

\begin{document}

\title{On mechanism design with expressive preferences: an aspect of the social choice of Brexit{\footnote{Declarations: we have no declarations to make with regard to funding or conflicts of interest/competing interests etc. We are very grateful to John Bone and Arunava Sen for several helpful suggestions. For discussions we also thank John Hey and John Suckling.}}}

\author{Anindya Bhattacharya{\footnote{Department of Economics and Related Studies, University of York, York, YO10 5DD, UK; anindya.bhattacharya@york.ac.uk\vspace{0.1cm}}}\hsp
Debapriya Sen{\footnote{Department of Economics, Toronto Metropoliton University, Toronto, ON, Canada; dprsen@gmail.com\vspace{0.2cm}}}}

\date{\today}

\maketitle

\begin{abstract}\noi We study some problems of collective choice when individuals can have expressive preferences, that is, where a decision-maker may care not only about the material benefit from choosing an action but also about some intrinsic morality of the action or whether the action conforms to some identity-marker of the decision-maker. We construct a simple framework for analyzing mechanism design problems with such preferences and present some results focussing on the phenomenon we call ``Brexit anomaly". The main findings are that while deterministic mechanisms are quite susceptible to Brexit anomaly, even with stringent domain restriction, random mechanisms assure more positive results.

\end{abstract}

\vspace{0.4cm} \noi {\it Keywords}: expressive preferences; mechanism; Brexit anomaly  

\vspace{0.4cm} \noi {\it JEL Classification}: D71; D72; D02

\vspace{0.4cm}

\newpage

\section{Introduction}

In contexts of decision-making, presence of ``expressive" preferences|where a decision-maker may care not only about the material benefit from choosing an action but also about some intrinsic morality of the action or whether the action conforms to some individual  identity-marker|and its implications have received some attention (see, e.g., Hillman 2010; Borah 2019 and the references therein). Perhaps the most dramatic contemporary example of such preferences at work is the ``Brexit referendum" of 2016: the collective decision about whether the United Kingdom (UK) would leave the European Union (EU) or not. Below we quote from Lynskey (2017):

\begin{quote}

\small{\ldots On the morning of 23 June 2016, Rosamund Shaw still wasn’t sure if she wanted Britain to leave the European Union. During the preceding weeks, she had been in turmoil. She absorbed a stream of negative stories about the EU in the Daily Mail, but wasn’t sure they were reliable\ldots In the voting booth, Shaw finally made her choice: she voted leave. ``{\em To be quite frank, I did not believe it would happen," she says. ``I thought I’d put in a protest vote. The impact of my stupidity!"} [Emphasis ours] As soon as Shaw saw the result the following morning, her heart sank. ``I was in shock," she remembers. ``Even though I voted leave, I thought, ‘Oh no! This is terrible!’\\
\ldots \\ 
Mark Olive is a 30-year-old serviceman who lives in the south of England. Before the referendum was announced, he had never thought much about the EU, so he tasked himself with reading as much information as he could from all sides. ``I was just getting a negative feeling about the EU, like it didn’t serve the interests of our country," he says. The main reason that he was undecided until referendum day was the tenor of the leave campaign \ldots

{\em Nonetheless, he went for leave. ``I went to sleep thinking that we weren’t going to leave and in the morning I was shocked."} [Emphasis ours]}

\end{quote}This example seems to have the following feature. Each individual in a society has to choose one of the two acts: $r$ (remain), $\ell$ (leave). There are two corresponding social states: $\hat{r}$ (society chooses to remain), $\hat{\ell}$ (society chooses to leave). The resulting social state is $\hat{r}$ if the majority chooses $r$ and it is $\hat{\ell}$ if the majority chooses $\ell.$ There are individuals who prefer to choose $r$ under some action choices by others. But if the probability of the event that an individual's choice would indeed change the social state from $\hat{r}$ to $\hat{\ell}$ is very small, some individuals prefers $\ell$ to $r.$ The action $\ell$ in this example above encapsulates the utility from ``moral protest" against the bureaucratic functionings of the EU.

The stories about Brexit quoted above show that outcomes from expressive behavior cannot be explained merely by invoking the presence of externality. The individuals did not just abstain (assuming that the UK would ``remain" irrespective of their voting); they actually incurred the cost of physically going to voting booths and casting their votes. So, the very act of making a ``protest vote" provided them with some payoff (within a supposed social state that the UK would ``remain"). 

Analyses of such behavior (which we formalize in the next section and name it ``Brexit anomaly") in the specific context of voting are there (see, e.g., Feddersen et al. 2009; Borah 2019 and the references therein). The focus of analysis for these works is to identify conditions, given {\em specific} voting protocols, under which voting expressively might emerge as an ``equilibrium'' behavior for at least a part of the voters.

We recognize that a central point for understanding expressive behavior is that such agents can be modeled as having preference over social states as well as actions. Thus, the presence of such behavior induces one to construct and look into a more general framework of the problems of implementation (especially with agents having private information).{\footnote{However, some works in complete information environments are there: e.g., Dutta and Sen (2012), Kimya (2017).}} 

As an example, suppose there is a finite set of voters and a set of two candidates $Z=\{1,2\}.$ Each voter can either abstain or vote for one of the candidates. Denote by $a^0$ the act of abstaining and $a^j$ the act of voting for candidate $j$ for $j=1,2.$  The identical set of actions of all voters is $A=\{a^0,a^1,a^2\}.$ Each voter has a preference ordering over $A \times Z.$ For instance, a specific voter $i$ may have a preference that has $(a^1,1) \succ_i (a^0,1) \succ_i (a^0,2) \succ_i (a^1,2),$ reflecting that voter $i$ would like to get  candidate $1$ elected, but in the scenario candidate $2$ is elected, does not want to get into the trouble of voting. 

Suppose that the preference ordering of any voter is its private information. Now assume that a planner has a multi-valued social choice function (i.e., a social choice correspondence) such that for every realized profile of preference orderings, the winner should be the one preferred by the majority and no more than 5\% of the voters would abstain. The issue might be that how would a design ``implement" this social choice correspondence of the planner. 

In this paper we do not address the general question outlined above fully but our work here is in that direction. Our goal in this paper is, first, to set up a {\em general} framework for analyzing collective choice in presence of expressive preferences and then, in that framework, to investigate axiomatic properties of {\em usual} mechanisms: i.e., which map a profile of actions chosen by the agents to a specific social state. In this paper we concentrate on the phenomenon of ``Brexit anomaly". 

We envisage a situation as follows. A planner knows that each agent in a society has a preference ordering over the set of all pairs of actions and social states. For each agent the actual realized preference ordering is its private information but  the planner knows the domain (that is, the set in which this ordering belongs). The goal of the planner is to design a mechanism (deterministic or random) which picks a social state (or a lottery over the set of social states in case of random mechanisms) for every profile of actions chosen by the individuals so that the mechanism avoids ``Brexit anomaly". That is, we seek to explore, as is usual from the perspective of classical mechanism design literature, what kinds of mechanisms (not merely, say, majority or plurality voting), if any, can avoid Brexit anomaly when the domains of individual preferences are sufficiently rich and what kinds of intuitive and acceptable domain restrictions help in avoiding such anomaly.\footnote{While the recent paper by Moldovanu and Rosar (2021) also deals with some issues related to the Brexit referendum, their focus of interest is quite different from ours.} 

First we find a straightforward characterization of the class of deterministic mechanisms which can avoid Brexit anomaly when preferences are unrestricted; this condition is quite stringent. Next we find that if the domain of preferences is sufficiently rich for each individual then no reasonable mechanism (the ones satisfying both unanimity and monotonicity) in the specific environment of one-shot voting can avoid Brexit anomaly. Further, we show that even with quite a stringent restriction like each preference orderings being separable in actions and social states (defined precisely in the following section), Brexit anomaly may be possible for well-known mechanisms. Therefore, we explore probabilistic aggregation. One interesting finding is that with a finite set of outcomes, if each individual's preference is strict then the sub-class of random mechanisms for which each action profile gives a probability distribution that assigns positive probability to every outcome avoids Brexit anomaly.

The paper is organized as follows. Section \ref{sec2} gives the initial framework and analyzes the case of deterministic mechanisms. Section \ref{sec3} explores the specific contexts of one-shot voting and queueing. The results for probabilistic mechanisms are given and discussed in Section \ref{sec4}. The final section contains some concluding remarks.

\section{A framework for aggregating expressive preferences and deterministic aggregation}

\label{sec2}

\hyperref[sec:sec2]{}

Let $N=\{1,\ldots,n\}$ be a finite set of $n$ agents. Each agent $i \in N$ has a set of actions $A_i.$ The elements of $A_i$ are denoted by $x_i,a_i$ etc. The set of all profiles of actions is $\Pi_{i=1}^n A_i$ and let $A_{-i}=\Pi_{j\neq i}A_j.$ The elements of $A_{-i}$ are denoted by $x_{-i},$ $a_{-i}$ etc. The profile of actions where $x_i\in A_i,$ $a_{-i}\in A_{-i}$ is denoted by $(x_i,a_{-i}).$   

Let $Z$ be the set of all outcomes (social states). Elements of $Z$ are denoted by $z,\tilde{z}$ etc. Each agent $i \in N$ has a  preference ordering $\succsim_i$ over $A_i \times Z.$  For $(x_i,z),(\tilde{x}_i,\tilde{z})\in A_i \times Z,$  ``$(x_i,z)\succsim_i(\tilde{x}_i,\tilde{z})$" stands for ``agent $i$ weakly prefers the action-outcome pair $(x_i,z)$ to the action-outcome pair $(\tilde{x}_i,\tilde{z})$". Agent $i$ {\it strictly prefers} $(x_i,z)$ to $(\tilde{x}_i,\tilde{z}),$ denoted by $(x_i,z)\succ_i(\tilde{x}_i,\tilde{z}),$ if $[(x_i,z)\succsim_i(\tilde{x}_i,\tilde{z})]$ and $[\mbox{not }(\tilde{x}_i,\tilde{z})\succsim_i (x_i,z)].$ Agent $i$ is {\it indifferent between} $(x_i,z)$ and $(\tilde{x}_i,\tilde{z}),$ denoted by $(x_i,z)\sim_i(\tilde{x}_i,\tilde{z}),$ if  $[(x_i,z)\succsim_i (\tilde{x}_i,\tilde{z})]$ and $[(\tilde{x}_i,\tilde{z})\succsim_i (x_i,z)].$

Let ${\cal R}_i$ be the set of all complete, transitive and reflexive preference orderings on $A_i \times Z.$ For each $i \in N,$ the {\em admissible} set of preference orderings on $A_i \times Z$ is denoted by ${\cal D}_i \subseteq {\cal R}_i.$ An {\em environment} ${\cal E}$ is a collection $\langle N, Z, {\{A_i\}_{i \in N}},{\{{\cal D}_i\}_{i \in N}}\rangle.$ Given an environment ${\cal E},$ an aggregation rule or mechanism $f: \Pi_{i=1}^n A_i \mapsto Z$ picks an outcome for every profile of actions. The outcome that $f$ picks for the profile $(x_i,a_{-i})$ is denoted by $f(x_i,a_{-i}).$ A first formalization of the anomaly in aggregation is as follows.

\begin{defn} \label{nba} \rm A mechanism $f$ satisfies ``Brexit anomaly" (BA) in an environment ${\cal E}$ if there exist $i \in N,$ distinct $r_i,\ell_i \in A_i,$ distinct $a_{-i},$ $b_{-i}\in A_{-i}$ and a preference ordering $\succsim_i \in {\cal D}_i$ such that

(i)    $f(r_i, a_{-i})=f(\ell_i, a_{-i}),$

(ii)  $(\ell_i, f(\ell_i, a_{-i})) \succ_i (r_i, f(r_i, a_{-i})),$

(iii)   $(r_i, f(r_i, b_{-i})) \succsim_i (x_i, f(x_i, b_{-i}))$ for all $x_i\in A_i$

\noi Otherwise, $f$ satisfies ``no Brexit anomaly" (NBA).

\end{defn}

To recast the specific example of Brexit into the formalism of Definition \ref{nba}, suppose there are $2m+1$ voters, where $m\geq 1.$ The set of actions for each voter $i$ is $A_i=\{0,r_i,\ell_i\}$ where $r_i$ represents ``to vote remain", $\ell_i$ ``to vote leave" and $0$ ``to abstain". The mechanism $f$ relevant to Brexit is: if the number of votes to leave is more than the number to remain, then the UK leaves the EU; otherwise it remains in the EU. Specific $a_{-i},b_{-i}$ that satisfy the conditions of Definition \ref{nba} are as follows. For an individual voter $i,$ $a_{-i}$ is a profile of votes where less than $m$ of the remaining voters vote to leave, while $b_{-i}$ is a profile where exactly $m$ of the remaining voters vote to leave.

When the number of remaining votes to leave is less than $m,$ the outcome $f$ will be ``remain" regardless of what voter $i$ chooses. Thus $f(r_i, a_{-i})=f(\ell_i, a_{-i}),$ as stated in condition (i) of Definition \ref{nba}. In this case voter $i$ prefers to record its protest by choosing $\ell_i,$ that is, $(\ell_i,f(\ell,a_{-i}))\succ_i (r_i,f(r_i,a_{-i})),$ as stated in condition (ii). 

When the number of remaining votes to leave is $m,$ the outcome will be ``remain" if $i$ chooses $r_i$ and ``leave" if $i$ chooses $\ell_i.$ In this case  voter $i$ prefers to choose $r_i$ to ensure the outcome is ``remain", that is, $(r_i,f(r_i,b_{-i}))\succ_i (\ell_i,f(\ell_i,b_{-i})).$ Condition (iii) of Definition \ref{nba} more generally requires that there is $b_{-i}$ at which $r_i$ is the most preferred action for $i.$   

When the domain ${\cal D}_i$ of preference orderings for each agent $i\in N$ is sufficiently rich, there is an elementary but complete characterization of the mechanisms which satisfy NBA. This necessary and sufficient condition turns out to be quite stringent.

\begin{prop} \label{prop1} {\it For an environment ${\cal E}$ where ${\cal D}_i={\cal R}_i$ for each $i\in N,$ a mechanism $f$ satisfies NBA if and only if the following holds for all $i\in N$ and all distinct $r_i,\ell_i\in A_i$:
$$f(r_i, a_{-i})=f(\ell_i, a_{-i})=z\mbox{ for some }a_{-i}\in A_{-i}\mbox{ and some }z\in Z$$
\be\label{con}\Rightarrow f(r_i, b_{-i})=f(\ell_i, b_{-i})=z\mbox{ for all }b_{-i}\in A_{-i}\ee}
\end{prop}\noi {\bf Proof} [If:] Suppose condition (\ref{con}) holds for $f.$ If $f$ does not satisfy NBA, then BA holds for $f.$ By (i) of Definition \ref{nba}, $\exists$ $i\in N,$ distinct $r_i,\ell_i\in A_i$ and $a_{-i}\in A_{-i}$ such that $f(r_i, a_{-i})=f(\ell_i, a_{-i})=z$ for some $z\in Z.$ Using this in (ii) of Definition \ref{nba}, we have $(\ell_i,z)\succ_i(r_i, z).$ Taking $x_i=\ell_i$ in (iii) of Definition \ref{nba}, there is $b_{-i}$ such that $(r_i,f(r_i,b_{-i}))\succsim_i(\ell_i,f(\ell_i,b_{-i})).$ By (\ref{con}), we have $f(r_i, b_{-i})=f(\ell_i, b_{-i})=z,$ so that $(r_i,z)\succsim_i(\ell_i,z),$ a contradiction since $(\ell_i,z)\succ_i(r_i, z).$ This shows that $f$ must satisfy NBA.

[Only if:] Suppose $f$ satisfies NBA. If condition (\ref{con}) does not hold, $\exists$ $i\in N,$ distinct $r_i,\ell_i\in A_i,$ distinct $a_{-i},b_{-i}\in A_{-i},$ $z\in Z$ such that $f(r_i, a_{-i})=f(\ell_i, a_{-i})=z$ and at least one of $f(r_i, b_{-i}),$ $f(\ell_i, b_{-i})$ does not equal $z.$ Thus either $(r_i,f(r_i,b_{-i}))\neq (r_i,z)$ or $(\ell_i,f(\ell_i,b_{- i}))\neq (\ell_i,z)$ (or both). Therefore $\exists$ $\succsim_i\in {\cal D}_i$ that has\footnote{Note that ${\cal D}_i={\cal R}_i,$ which is the set of all complete and transitive preference orderings on $A_i\times Z.$} 
$$\mbox{ (I) }(\ell_i,z)\succ_i (r_i,z),\mbox{ (II) }(r_i,f(r_i,b_{-i}))\succsim_i (\ell_i,f(\ell_i,b_{-i})),$$\be\label{m0}\mbox{ and in addition (III) }(r_i,f(r_i,b_{-i}))\succsim_i (x_i,f(x_i,b_{-i}))\mbox{ for all }x_i\in A_i \setminus \{r_i,\ell_i\}\ee (II)-(III) imply (iii) of Definition \ref{nba} and because $f(r_i, a_{-i})=f(\ell_i, a_{-i})=z,$ (I) implies (i)-(ii) of Defintion \ref{nba}. Thus $f$ satisfies BA, a contradiction. This shows condition (\ref{con}) must hold. \qed 

\begin{rem}\rm \label{rem0} Let ${\cal P}_i$ be the set of all strict preference orderings on $A_i\times Z.$ Replacing $\succsim_i$ by $\succ_i$ in (\ref{m0}), it follows that we can find a preference ordering in ${\cal P}_i$ that satsifies (\ref{m0}). This shows that the conclusion of Proposition \ref{prop1} continues to hold with restricted domains ${\cal D}_i={\cal P}_i$ as well as ${\cal D}_i={\cal R}_i\setminus {\cal P}_i.$
\end{rem}

\begin{rem}\rm \label{rem01} If every agent $i$ has a {\it classical} preference in that it depends only on the outcomes, then for all $r_i,\ell_i\in A_i$ and $z\in Z,$ we have $(r_i,z)\sim_i (\ell_i,z).$ In that case, conditions (i),(ii) of Definition \ref{nba} can never hold together and therefore NBA holds. For example, consider a Cournot duopoly with two firms $1,2$ who  have identical constant marginal cost $c$ and can choose any non-negative quantities, so $A_1=A_2=\mathbb{R}_+.$ Firms face inverse demand $p:\mathbb{R}_+\mapsto \mathbb{R}_+,$ so for any $(x_1,x_2)\in A_1\times A_2,$ the price is $p(x_1+x_2)$ and the outcome is $$f(x_1,x_2)=(p(x_1+x_2),\pi_1(x_1,x_2),\pi_2(x_1,x_2))$$where $\pi_i(x_1,x_2)=[p(x_1+x_2)-c]x_i$ is the profit of firm $i$ at $(x_1,x_2).$  Note that $f(r_1,a_2)=f(\ell_1,a_2)$ implies $\pi_1(r_1,a_2)=\pi_1(\ell_1,a_2).$ If firm $1$ cares only about its own profit, whenever $f(r_1,a_2)=f(\ell_1,a_2),$ we must have $(r_1,f(r_1,a_2))\sim_1 (\ell_1,f(\ell_1,a_2)),$ so both conditions (i),(ii) of Definition \ref{nba} cannot hold for firm $1.$ Similarly if firm $2$ cares only about its own profit, both conditions (i),(ii) cannot hold for firm $2.$ Thus in a standard Cournot duopoly, where the preference of each firm depends only on the outcome, BA does not hold.  The same conclusion holds for an oligopoly. More generally BA does not hold for a game where the payoff of each player depends on action profiles and each player cares only about its own payoff.
\end{rem}

\begin{rem}\rm \label{rem1} A literature on analyses of aggregation where the set of social states is some subset of a multidimensional space exists (see, e.g., Austen-Smith and Banks 1999; Le Breton and Sen 1999). However, the aggregation problem which we look into is not quite similar to that either. The framework of Suzumura and Xu (2001), where desirability of an act of choice changes as  
the opportunity set (from which the choice is to be made) changes has a bit of {\em apparent} similarity with our framework, but the underlying substantive contents of the two are quite different. Further, as the example in the Section 3 illustrates, our framework is a natural setting for analyzing several problems of mechanism design with behavioral players (e.g., a number of scenarios in Kimya 2017 are amenable to analysis from within our framework).

Further note that while in the present framework of ours agents' preferences are {\em ordinal}, at least one alternative {\em cardinal} framework for approaching the same issue exists. Alternatively, one could take a full-fledged {\em cardinal} framework with costly actions (as, e.g., in Kartik, 2009) with private information, but common knowledge of the information structures of the agents. While that can be an item for further research, our framework is evidently simpler and demands much less structure.

\end{rem}

\section{Some specific contexts}

\label{sec3}

\hyperref[sec:sec3]{}

We explore Brexit anomaly for certain specific contexts of aggretation. 

\subsection{Simple voting}

In the environment of simple voting we find that if the domain of preferences is sufficiently rich for each voter, then no reasonable mechanism can avoid BA. Consider an environment ${\cal E}=$ $\langle N, Z, {\{A_i\}_{i \in N}},{\{{\cal D}_i\}_{i \in N}}\rangle$ where $N=\{1,\ldots,n\}$ is the set of voters, $Z=\{1,\ldots,m\}$ the set of candidates with $n,m\geq 2$ and ${\cal D}_i$ the set of admissible preference orderings on $A_i \times Z.$ Each voter $i \in N$ has the identical set of actions $A_i=\{0\}\cup Z$ where action $0$ is ``to abstain" and for $z\in Z,$ action $z$ stands for ``vote for candidate $z$". A mechanism $f:\Pi_{i\in N}A_i\mapsto Z$ in this environment picks one of the candidates as the ``winner" for each profile of votes.

A mechanism $f$ satisfies {\em unanimity} if whenever every $i\in N$ votes for the same candidate $z,$ $f$ picks $z.$ It satisfies {\em monotonicity} if for any $i\in N,$ $x_i\in A_i,$ $a_{-i}\in A_{-i},$ $z\in Z$: $f(x_i,a_{-i})=z$ implies $f(z,a_{-i})=z.$

\begin{prop} \label{prop2} {\it Consider an environment of simple voting with at least two voters and candidates. Suppose either ${\cal D}_i={\cal R}_i$ for all $i\in N$ or ${\cal D}_i={\cal P}_i$ for $i\in N$ or ${\cal D}_i={\cal R}_i\setminus {\cal P}_i$ for all $i\in N.$ Then  any mechanism $f$ satisfying unanimity and monotonicity suffers from Brexit anomaly $(BA)$.}

\end{prop}\noi {\bf Proof} We show that condition (\ref{con}) fails to hold for any $f$ satisfying unanimity and monotonicity. Then it follows by Proposition \ref{prop1} that BA must hold for any such $f.$ 

To assess condition (\ref{con}) for $f,$ consider any $i\in N,$ $r_i\in A_i=\{0\}\cup Z$ with $r_i\in Z$ (that is, $r_i$ is not the action to abstain), $a_{-i}\in A_{-i}$ and let $f(r_i,a_{-i})=z.$ Suppose $z\neq r_i.$ Taking $\ell_i=z,$ by monotoncity of $f$ we have $f(\ell_i,a_{-i})=f(z,a_{-i})=z.$ Observe that $a_{-i}$ must have at least one voter $j$ with $a_j\neq r_i$ (otherwise unanimity will imply $f(r_i,a_{-i})=r_i,$ contradicting $z\neq r_i$). Let $b_{-i}\in A_{-i}$ be the profile with $b_j=r_i$ for all $j\neq i.$ Then by unanimity, $f(r_i,b_{-i})=r_i.$ Thus we have distinct $r_i,\ell_i\in A_i,$ distinct $a_{-i},b_{-i}\in A_{-i}$ such that $f(r_i,a_{-i})=f(\ell_i,a_{-i})=z$ but $f(r_i,b_{-i})=r_i\neq z,$ violating condition (\ref{con}). 

Thus for condition (\ref{con}) to hold for $f,$ for every $i\in N,$ $r_i\in Z,$ $a_{-i}\in A_{-i},$ we must have $f(r_i,a_{-i})=r_i,$ that is, whenever $i$ votes for a specific candidate, $f$ picks that candidate regardless of how others vote. But given that there are at least two voters and candidates, this cannot be true for all $i\in N.$ This shows condition (\ref{con}) does not hold for $f$ and therefore $f$ satisfies BA. \qed

\begin{rem}\label{rem2} Propositions \ref{prop1} and \ref{prop2} illustrate the following facts. First, BA can happen with respect to very few individuals in the population (in Proposition \ref{prop2} it is relevant, possibly, for two individuals only!) and thus, may turn out to be quite rare in the real world. But, secondly, with an unrestricted (or sufficiently rich) domain of preferences, deterministic mechanisms are quite amenable to BA; avoiding such anomaly with deterministic aggregation seems quite difficult. In such scenarios one usual route is to explore the implications of domain-restrictions. Since one source of this anomaly arising is that persons have preferences over ``action-social state" pairs, an initial expectation might be that if the preference orderings satisfy ``separability" (defined precisely in the next section) in actions and social states then BA can be avoided. However, in the next section we show that this is not necessarily true.
\end{rem}

\subsection{The queueing problem}

Consider another specific example of mechanism design: the queueing problem (see, e.g., Mitra 2001). There are two patients $1,2,$ who visit a public clinic for a treatment. The treatment from the clinic gives a common benefit $\ol{u}>0$ to each patient. The clinic can see only one patient at a time and patients have to wait in a queue. The person placed first in the queue has no waiting time and incurs no waiting cost, but the person placed in the second position waits for one unit of time and incurs a waiting cost. For patient $i\in N= \{1,2\},$ the waiting cost for every unit of time is $\theta_i\in [0,1],$ which is private information to that patient. 

The clinic arranges the two patients in a queue (equivalently, assigns each patient a waiting time of either zero or one) and can receive transfers from the patients. In this example, an outcome or a social state is given by $(w,t)\equiv ((w_1,w_2),(t_1, t_2)),$ where $w_i\in\{0,1\}$ is the waiting time for patient  $i$ and $t_i\in \mathbb{R}$  the transfer (positive, negative or zero) that $i$ makes to the clinic. The set of all outcomes is 
\be\label{bz1}Z=\{(w,t)|w_1,w_2\in \{0,1\},w_1+w_2=1,t_1,t_2\in \mathbb{R}\}\ee 

The clinic asks each patient to report its unit waiting cost. Patient $i$ can report any $x_i\in [0,1],$ so the set of actions for $i$ is $A_i=[0,1].$ The environment for this queueing problem is ${\cal E}=\langle N, Z, {\{A_i\}_{i \in N}},{\{{\cal D}_i\}_{i \in N}}\rangle,$ where $N=\{1,2\},$ $A_i=[0,1],$ $Z$ is given by (\ref{bz1}) and ${\cal D}_i$ is the set of all admissible preference orderings over $A_i\times Z.$ A mechanism is a function $f:[0,1]\times [0,1]\rightarrow Z,$ that is, for every pair of reports $(x_1,x_2),$ a mechansim $f$ assigns a waiting time $w^f_i(x_1,x_2)$ and a transfer payment $t^f_i(x_1,x_2)$ for each $i\in N.$ 

\subsubsection{The Groves mechanism}

The Groves mechanism for this queuing problem is follows.

\begin{enumerate}[(i)]

\item Each patient $i\in N=\{1,2\}$ reports its unit waiting cost $x_i\in [0,1].$

\item If $x_1 > x_2$: patient $1$ is assigned waiting time zero and makes transfer $x_2$ to the clinic; patient $2$ is assigned waiting time one and makes no transfer. 

\item If $x_1 \leq x_2$: patient $2$ is assigned waiting time zero and makes transfer $x_1$ to the clinic; patient $1$ is assigned waiting time one and makes no transfer. 

\end{enumerate}Thus, for every pair of reports $(x_1,x_2),$ the waiting time $w^f_i(x_1,x_2)$ and the transfer payment $t^f_i(x_1,x_2)$ assigned to each $i\in N$ by the Groves mechanism $f:[0,1]\times [0,1]\rightarrow Z$ has
$$w^f_1(x_1,x_2)=0,t^f_1(x_1,x_2)=x_2,w^f_2(x_1,x_2)=1,t^f_2(x_1,x_2)=0\mbox{ if }x_1> x_2$$
\be\label{bp1}w^f_1(x_1,x_2)=1,t^f_1(x_1,x_2)=0,w^f_2(x_1,x_2)=0,t^f_2(x_1,x_2)=x_1\mbox{ if }x_1\leq x_2\ee  

\begin{prop}\label{prop3} {\it Consider the environment of the queueing problem. Suppose either ${\cal D}_i={\cal R}_i$ for all $i\in N$ or ${\cal D}_i={\cal P}_i$ for $i\in N$ or ${\cal D}_i={\cal R}_i\setminus {\cal P}_i$ for all $i\in N.$ The Groves mechanism $f$ in this environment suffers from Brexit anomaly $(BA)$.}
\end{prop}\noi {\bf Proof} Take $r_1,\ell_1,a_2,b_2\in [0,1]$ such that $a_2<r_1<\ell_1<b_2.$ By (\ref{bp1}), $f(r_1,a_2)=f(\ell_1,a_2)$ but $t^f_2(r_1,b_2)=r_1<t^f_2(\ell_1,b_2)=\ell_1,$ so $f(r_1,b_2)\neq f(\ell_1,b_2).$ This shows condition (\ref{con}) of Proposition \ref{prop1} fails to hold for $f$ and therefore $f$ satisfies BA. \qed \\

Note from Proposition \ref{prop3} that even with very stringent domain restrictions, Brexit anomaly may appear even for a simple (perhaps the simplest in this context) Groves mechanism. The next section presents a constructive proof of BA for the Groves mechanism $f$ given in (\ref{bp1}) by showing that a specific expressive preference $\succsim_i$ satisfies (i)-(iii) of Definition \ref{nba}. 

\subsubsection{Specific expressive preferences} At any outcome $(w,t)\in Z,$ the {\it material payoff} of patient $i\in N,$ denoted by $u_i:Z\rightarrow \mathbb{R},$ is given by the benefit $\ol{u}$ net of the patient's waiting cost and the transfer payments. If patient $i$ has waiting time $w_i$ and makes transfer $t_i$ to the clinic, its waiting cost is $\theta_iw_i$ and so its material payoff at $(w,t)$ is 
\be\label{mp}u_i(w,t)=\ol{u}-\theta_iw_i-t_i\ee 

Patient $1$ has the following expressive preference $\succsim_1$ over  $A_1 \times Z.$ For any $x_1,\tilde{x}_1\in A_1$ and $(w,t),(\tilde{w},\tilde{t})\in Z$:

\begin{enumerate}[(a)]

\item If $u_1(w,t)>u_1(\tilde{w},\tilde{t}),$ then $(x_1,(w,t))\succ_1 (\tilde{x}_1,(\tilde{w},\tilde{t})).$ 

\item If $u_1(w,t)=u_1(\tilde{w},\tilde{t})$ and $t_1+t_2>\tilde{t}_1+\tilde{t}_2,$ then  $(x_1,(w,t))\succ_1 (\tilde{x}_1,(\tilde{w},\tilde{t})).$ 

\item If $u_1(w,t)=u_1(\tilde{w},\tilde{t}),$ $t_1+t_2=\tilde{t}_1+\tilde{t}_2$ and $|\theta_1-x_1| < |\theta_1-\tilde{x}_1|,$  then $(x_1,(w,t))\succ_1 (\tilde{x}_1,(\tilde{w},\tilde{t})).$ 

\item If $u_1(w,t)=u_1(\tilde{w},\tilde{t}),$ $t_1+t_2=\tilde{t}_1+\tilde{t}_2$ and $|\theta_1-x_1| = |\theta_1-\tilde{x}_1|,$  then $(x_1,(w,t))\sim_1 (\tilde{x}_1,(\tilde{w},\tilde{t})).$

\end{enumerate}In words, the preference of patient $1$ over action-outcome pairs has a lexicographic nature with a hierarchy of importance of three factors: (i) material payoff, (ii) revenue (that is, total transfer payments) to the clinic and (iii) partial honesty (see Dutta and Sen, 2012). Observe that both the material payoff and revenue to the clinic depend only on the outcomes. 

Property (a) above says for two action-outcome pairs with different material payoffs, patient $1$ prefers the one that gives a higher material payoff. Property (b) says for two alternatives with the same material payoff, $1$ prefers the one that gives higher revenue  to the clinic. Property (c) says for two alternatives with the same material payoff and revenue, $1$ prefers the one in which its report is closer to its true unit waiting cost $\theta_1.$ Finally property (d) says that $1$ is indifferent between two alternatives that are same with regard to all of these three factors.

Patient $2$ has a {\it classical} preference in that $2$ only cares about its material payoff given in (\ref{mp}). Formally the expressive preference $\succsim_2$ of $2$ over  $A_2\times Z$ (action-outcome pairs) is as follows. For any $x_1,\tilde{x}_1\in A_2$ and $(w,t),(\tilde{w},\tilde{t})\in Z$: if $u_2(w,t)>u_2(\tilde{w},\tilde{t}),$ then $(x_2,(w,t))\succ_1 (\tilde{x}_2,(\tilde{w},\tilde{t}))$ and if $u_2(w,t)=u_2(\tilde{w},\tilde{t}),$ then $(x_2,(w,t))\sim_2 (\tilde{x}_2,(\tilde{w},\tilde{t})).$

\vspace{0.1cm} Recall the notion of separable preference orderings (see, e.g., Fishburn, 1970, p.43).

\begin{defn} \label{sep}\rm  A preference ordering $\succsim_i$ on $A_i \times Z$ is {\it separable} if for all $x_i,\tilde{x}_i\in A_i$ and $(w,t),(\tilde{w},\tilde{t})\in Z$:

 (i) $(x_i,(w,t)) \succsim_i (x_i,(\tilde{w},\tilde{t}))$ implies $(\tilde{x}_i,(w,t)) \succsim_i (\tilde{x}_i,(\tilde{w},\tilde{t})).$

 (ii) $(x_i,(w,t)) \succsim_i (\tilde{x}_i,(w,t))$ implies $(x_i,(\tilde{w},\tilde{t})) \succsim_i (\tilde{x}_i,(\tilde{w},\tilde{t})).$
\end{defn}

\begin{ob}\label{ob1} {\it Preferences $\succsim_1,\succsim_2$ are both separable.}

\end{ob}\noi {\bf Proof} See the Appendix. \qed

\begin{prop} \label{prop4} {\it For any $\theta_1\in (0,1),$ there are distinct $r_1,\ell_1\in [0,1],$ distinct $a_2,b_2\in [0,1]$ such that} (i)-(iii) {\it of Definition $\ref{nba}$ holds for $\succsim_1$ under the Groves mechanism $f$ given in $(\ref{bp1}),$ that is}, 
(i)  $f(r_1, a_2)=f(\ell_1, a_2),$ (ii)  $(\ell_1, f(\ell_1, a_2)) \succ_1 (r_1, f(r_1, a_2))$
{\it and} (iii)  {\it $(r_1, f(r_1, b_2)) \succsim_1 (x_1, f(x_1, b_2))$ for all $x_1\in [0,1].$}
\end{prop}
\noi {\bf Proof} First we recall from (\ref{mp}) that the material payoff of each patient is the benefit $\ol{u}$ net of its waiting cost and the transfer payments. Using (\ref{mp}), the  material payoffs of patients $1,2$ under the Groves mechanism $f$ of  (\ref{bp1}) is 
$$u^f_1(x_1,x_2)=\ol{u}-x_2,u^f_2(x_1,x_2)=\ol{u}-\theta_2\mbox{ if }x_1\geq x_2$$
\be\label{mu1}u^f_1(x_1,x_2)=\ol{u}-\theta_1,u^f_2(x_1,x_2)=\ol{u}-x_1\mbox{ if }x_1<x_2\ee 

Let $\theta_1\in (0,1)$ and $r_1,\ell_1,a_2,b_2\in [0,1]$ be such that $a_2<\ell_1=\theta_1<r_1=b_2.$ Since both $r_1,\ell_1$ are more than $a_2,$ by  (\ref{bp1}), $$w^f_1(r_1,a_2)=w^f_1(\ell_1,a_2)=0,t^f_1(\ell_1,a_2)=t^f_1(r_1,a_2)=a_2,$$
$$w^f_2(r_1,a_2)=w^f_2(\ell_1,a_2)=1,t^f_2(r_1,a_2)=t^f_2(\ell_1,a_2)=0$$This shows the $f$ gives the same outcome for $(r_1,a_2),$ $(\ell_1,a_2),$ that is, $f(r_1,a_2)=f(\ell_1,a_2),$ so (i) of Definition \ref{nba} holds.

By (\ref{mu1}), the material payoffs and the revenue to the clinic are same at $f(r_1,a_2),$ $f(\ell_1,a_2)$:
$$u^f_1(r_1,a_2)=u^f_1(\ell_1,a_2)=\ol{u}-a_2,u^f_2(r_1,a_2)=u^f_2(\ell_1,a_2)=\ol{u}-\theta_2\mbox{ and }
$$$$t^f_1(r_1,a_2)+t^f_2(r_1,a_2)=t^f_1(\ell_1,a_2)+t^f_2(\ell_1,a_2)=a_2$$
Noting that $|\ell_1-\theta_1|=0<|r_1-\theta_1|,$ from property (c) of $\succsim_1$ we conclude that $(\ell_1,f(\ell_1,a_2))\succ_1 (r_1,f(r_1,a_2)),$ so (ii) of Definition \ref{nba} also holds.

Note from (\ref{bp1}) that for any $x_1\leq b_2$: 
$w^f_1(x_1,b_2)=1,$ $t^f_1(x_1,b_2)=0,$ $w^f_2(x_1,b_2)=0,$ $t^f_2(x_1,b_2)=x_1.$ Then by (\ref{mu1}) 
\be\label{p4}u^f_1(x_1,b_2)=\ol{u}-\theta_1,t^f_1(x_1,b_2)+t^f_2(x_1,b_2)=x_1\mbox{ for all }x_1\leq b_2\ee
This shows $f(x_1,b_2)$ gives the same material payoff $\ol{u}-\theta_1$ to patient $1$ for all $x_1\leq b_2,$ but the revenue to the clinic (total transfer payments $x_1$) is maximum when $x_1$ equals $b_2.$ Recall $r_1=b_2.$ Then by property (b) of $\succsim_1,$ we conclude
$$(r_1,f(r_1,b_2))\succsim_1 (x_1,f(x_1,b_2))\mbox{ for all }x_1\leq b_2$$
Again by (\ref{bp1}), for any $x_1> b_2$: 
$w^f_1(x_1,b_2)=0,$ $t^f_1(x_1,b_2)=b_2$ and so $u^f_1(x_1,b_2)=\ol{u}-b_2.$ 
Because $\theta_1<r_1=b_2,$ by (\ref{p4}), $u^f_1(r_1,b_2)=\ol{u}-\theta_1>\ol{u}-b_2.$ This shows $u^f_1(r_1,b_2)>u^f_1(x_1,b_2)$ for all $x_1>b_2.$ Applying property (a) of $\succsim_1,$ we conclude
$$(r_1,f(r_1,b_2))\succ_1 (x_1,f(x_1,b_2))\mbox{ for all }x_1>b_2$$
Thus $(r_1,f(r_1,b_2))\succsim_1 (x_1,f(x_1,b_2))$ for all $x_1\in [0,1],$ so (iii) of Definition \ref{nba} holds. \qed 

\section{Probabilistic aggregation: some possibility results}

\label{sec4}

\hyperref[sec:sec4]{}

\noi Since, as we saw in the previous section, avoiding Brexit anomaly seems very difficult with deterministic mechanisms, we explore in this section what happens with random mechanisms.

For studying probabilistic aggregation, we assume that the outcome-set $Z$ is finite containing at least two elements. Let $\Delta(Z)$ be the set of all probability distributions on $Z.$ For any probability distribution $g\in \Delta(Z),$ $g(z)$ is the probability with which outcome $z$ is chosen under $g$, so $g(z)\geq 0$ for all $z\in Z$ and $\sum_{z\in Z}g(z)=1.$ 

\begin{defn} \label{tmix} \rm A probability distribution $g\in \Delta(Z)$ is {\it totally mixed} if it assigns positive probability to each outcome, that is, $g(z)>0$ for all $z\in Z.$
\end{defn}

\subsection{First order stochastic dominance}

With probabilistic outcomes, we use, as is very common in the relevant literature, the concept of {\it first order stochsatic dominance} (FSD) for preferences over action-outcome pairs. As before consider an agent $i$ who has a (complete, transitive and reflexive) preference relation $\succsim_i$ on $A_i\times Z.$ Denote by $[r_i,g]$ the situation where agent $i$ chooses action $r_i$ and the probability distribution on outcomes is $g.$ Under $[r_i,g],$ agent $i$ obtains (i) action-outcome pair $(r_i,z)$ with probability $g(z)$ and (ii) action-outcome pair $(x_i,z)$ with probability zero for any $x_i\neq r_i.$

For any action-outcome pair $(x_i,z),$ denote by $\phi^{[r_i,g]}_i(x_i,z)$ the probability with which $i$ obtains an action-outcome pair which it weakly prefers to $(x_i,z)$ under $[r_i,g]$:  
\be\label{fsd1}\phi^{[r_i,g]}_i(x_i,z):=\sum_{\{\tilde{z}\in Z|(r_i,\tilde{z})\succsim_i (x_i,z)\}}g(\tilde{z})\ee  

\begin{defn} \label{fsd0} \rm Consider any agent $i$ who has a preference relation $\succsim_i$ on $A_i \times Z.$ For actions $r_i,\ell_i\in A_i$ and probability distributions $g,h\in \Delta(Z),$ we say $[r_i,g]$ {\it first order stochastically dominates $[\ell_i,h]$ for agent $i$} (denoted by ``$[r_i,g]$ FSD$_i$ $[\ell_i,h]$") if 
$$\phi^{[r_i,g]}_i(x_i,z)\geq \phi^{[\ell_i,h]}_i(x_i,z)\mbox{ for all }(x_i,z)\in A_i\times Z,$$ with strict inequality for at least one $(x_i,z).$
\end{defn}In words, $[r_i,g]$ FSD$_i$ $[\ell_i,h]$ if for any $(x_i,z),$ the probability of receiving action-outcome pairs that agent $i$ prefers more than $(x_i,z)$ is higher under $[r_i,g]$ compared to $[\ell_i,h].$

\begin{lem} \label{lemmix}{\it Suppose the set of outcomes $Z$ is finite with at least two elements. Consider any agent $i$ who has a strict preference relation $\succsim_i$ on $A_i \times Z.$ For distnct actions $r_i,\ell_i\in A_i,$ at least one of the statements}  (S1), (S2) {\it holds.}
{\small{$$\mbox{(S1) }\mbox{{\it for any totally mixed }}g,h\in \Delta(Z), \mbox{{\it neither} }``[\ell_i,g] \mbox{ FSD}_i \mbox{ }[r_i,h]"\mbox{ {\it nor} }``[\ell_i,h] \mbox{ FSD}_i \mbox{ }[r_i,g]"$$}}
{\small{$$\mbox{(S2) }\mbox{{\it for any totally mixed }}g,h\in \Delta(Z), \mbox{{\it neither} }
``[r_i,g] \mbox{ FSD}_i \mbox{ }[\ell_i,h]"\mbox{ {\it nor} }``[r_i,h] \mbox{ FSD}_i \mbox{ }[\ell_i,g]"$$ }}
\end{lem}\noi {\bf Proof} Since $Z$ is finite and $\succ_i$ is a strict preference ordering, for every $(x_i,z)\in A_i\times Z,$ $\exists$ a unique $z^*(x_i)\in Z$ such that $(x_i,z^*(x_i))\succ_i (x_i,z)$ for all $z\neq z^*(x_i).$ Consider any two distinct $r_i,\ell_i\in A_i.$ Because $\succ_i$ is a strict preference ordering, we have
\be\label{ot}\mbox{either (I) }(r_i,z^*(r_i))\succ_i (\ell_i,z^*(\ell_i))\mbox{ or (II) }(\ell_i,z^*(\ell_i))\succ_i (r_i,z^*(r_i))\ee First suppose (I) of (\ref{ot}) holds. Because $(r_i,z^*(r_i))\sim_i (r_i,z^*(r_i))$ and $(r_i,z^*(r_i))\succ_i (r_i,z)$ for all $z\neq z^*(r_i),$ we have $\{\tilde{z}\in Z|(r_i,\tilde{z})\succsim_i (r_i,z^*(r_i))\}=\{z^*(r_i)\}.$ Then for any totally mixed $g,h\in \Delta(Z),$  
$$\phi^{[r_i,h]}_i(r_i,z^*(r_i))=\sum_{\{\tilde{z}\in Z|(r_i,\tilde{z})\succsim_i (r_i,z^*(r_i))\}}h(\tilde{z})=h(z^*(r_i))>0$$
\be\label{ot1}\phi^{[r_i,g]}_i(r_i,z^*(r_i))=\sum_{\{\tilde{z}\in Z|(r_i,\tilde{z})\succsim_i (r_i,z^*(r_i))\}}g(\tilde{z})=g(z^*(r_i))>0\ee Next observe that under $[\ell_i,g],$ agent $i$ obtains (i) an action-outcome pair $(\ell_i,z)$ with probability $g(z)$ and (ii) an action-outcome pair $(x_i,z)$ with probability zero for any $x_i\neq \ell_i.$ Similarly under $[\ell_i,h],$ agent $i$ obtains (i) an action-outcome pair $(\ell_i,z)$ with probability $h(z)$ and (ii) an action-outcome pair $(x_i,z)$ with probability zero for any $x_i\neq \ell_i.$ Therefore when (I) of (\ref{ot}) holds, we have $\{\tilde{z}\in Z|(\ell_i,\tilde{z})\succsim_i (r_i,z^*(r_i))\}=\emptyset,$ so under both $[\ell_i,g]$ and $[\ell_i,h],$ agent $i$ never obtains an action-outcome pair $(x_i,z)$ with $(x_i,z)\succsim_i (r_i,z^*(r_i)).$ Hence  
$$\phi^{[\ell_i,g]}_i(r_i,z^*(r_i))=\sum_{\{\tilde{z}\in Z|(\ell_i,\tilde{z})\succsim_i (r_i,z^*(r_i))\}}g(\tilde{z})=0$$
\be\label{ot2}\phi^{[\ell_i,h]}_i(r_i,z^*(r_i))=\sum_{\{\tilde{z}\in Z|(\ell_i,\tilde{z})\succsim_i (r_i,z^*(r_i))\}}h(\tilde{z})=0\ee Using Definition \ref{fsd0}, the first equations of (\ref{ot1})-(\ref{ot2}) imply ``$[\ell_i,g]$ FSD$_i$ $[r_i,h]$" does not hold and the last  equations of (\ref{ot1})-(\ref{ot2}) imply ``$[\ell_i,h]$ FSD$_i$ $[r_i,g]$" does not hold. This shows when (I) of (\ref{ot}) holds, then (S1) holds. Similar reasoning shows when (II) of (\ref{ot}) holds, then (S2) holds. This completes the proof of the result. \qed 

\subsection{Probabilistic aggregation mechanisms}

Consider an environment where the set of outcomes $Z$ is finite with at least two elements,  each agent $i\in N$ has set of actions $A_i$ and preference relation $\succsim_i$ on $A_i\times Z.$  A  probabilistic mechanism $f: {\Pi}_{i \in N} A_i \mapsto \Delta (Z)$ associates a probability distribution on $Z$ to each profile of actions $x\in {\Pi}_{i \in N} A_i.$ We use the notion of FSD from Definition \ref{fsd0} to extend Defintion \ref{nba} to probabilistic mechanisms. 

\begin{defn} \label{nbap} \rm A probabilistic mechanism $f: {\Pi}_{i \in N} A_i \mapsto \Delta (Z)$ satisfies ``Brexit anomaly" (BA) in an environment ${\cal E}$ if there exist $i \in N,$ distinct $r_i,\ell_i \in A_i,$ distinct $a_{-i},$ $b_{-i}\in A_{-i}$ and a preference ordering $\succsim_i \in {\cal D}_i$ such that

(i)    $f(r_i, a_{-i})=f(\ell_i, a_{-i}),$

(ii)  $[\ell_i, f(\ell_i, a_{-i})]$ FSD$_i$ $[r_i, f(r_i, a_{-i})],$

(iii)   $[r_i, f(r_i, b_{-i})]$ FSD$_i$ $[x_i, f(x_i, b_{-i})]$ for all $x_i\in A_i.$

\noi Otherwise, $f$ satisfies ``no Brexit anomaly" (NBA).

\end{defn}

It is immediate that with probabilistic mechanisms avoiding BA might be quite trivial. Take, for example, the familiar probabilistic mechanism used extensively in probabilistic voting: the ``relative frequency" mechanism (defined below). Consider a simpler version of the environment that we studied in subsection 3.1 above. Take the environment ${\cal E}=$ $\langle N, Z, {\{A_i\}_{i \in N}},{\{{\cal D}_i\}_{i \in N}}\rangle$ where $N=\{1,\ldots,n\}$ is the set of voters, $Z=\{1,\ldots,m\}$ the set of candidates with $n,m\geq 2$ and ${\cal D}_i$ the set of admissible preference orderings on $A_i \times Z.$ Each voter $i \in N$ has the identical set of actions $A_i=Z$ where, for $z\in Z,$ action $z$ stands for ``vote for candidate $z$". Then the ``relative frequency" mechanism $R$ is defined as follows. Given any profile of actions $x,$ for any action $z \in Z,$ $R(z),$ the probability with which candidate $z$ is socially chosen, is given by $r_z /n$ where $r_z$ is the number of times $z$ occurs in the action profile $x.$ Then it is obvious that for the mechanism $R$ the Condition (i) in Definition \ref{nbap} is never satisfied and thus, $R$ satisfies NBA in such environments of simple voting.

However, in the Proposition \ref{prop1p} below we identify a rich class of mechanisms obeying NBA in {\em{general}} environments for which ${\cal D}_i={\cal P}_i$ for each $i \in N.$

For a probabilistic mechansim $f,$ denote by $f(x;z)$ the probability with which outcome $z$ is chosen under $f$ when the profile of actions is $x.$ Thus for every profile of actions $x,$ we have $f(x;z)\geq 0$ for all $z\in Z$ and $\sum_{z\in Z}f(x;z)=1.$ 

\begin{defn} \label{mix} \rm A probabilistic mechanism $f$ is {\it completely mixed} if for every profile of actions, it assigns positive probability to each outcome, that is, for every profile $x,$ $f(x;z)>0$ for all $z\in Z.$ 
\end{defn}

\begin{prop} \label{prop1p} {\it For an environment ${\cal E}$ where the set of outcomes $Z$ is finite with at least two elements and for each $i\in N,$ the domain of preference orderings is ${\cal P}_i$ $($the set of all strict preference orderings on $A_i\times Z).$ If a  probabilsitic mechanism $f$ is completely mixed, then it satisfies NBA.}
\end{prop}\noi {\bf Proof} Consider a probabilistic mechanism $f$ that is completely mixed. Consider any $i\in N$ and $\succ_i\in {\cal P}_i.$ Consider any two distinct $r_i,\ell_i\in A_i$ and $a_{-i},b_{-i}\in A_{-i}$  (if there does not exist two distinct elements in either $A_i$ or $A_{-i},$ then clearly the conditions of Definition \ref{nbap} do not hold for $i$). 

Since $f$ is completely mixed, $f(\ell_i, a_{-i}),$ $f(r_i, a_{-i}),$ $f(r_i, b_{-i}),$ $f(\ell_i, b_{-i})$ are all totally mixed probability distributions on $\Delta(Z).$ As $\succ_i$ is a strict preference ordering, we can apply the result of Lemma \ref{lemmix}. Observe that if (ii) of Definition \ref{nbap} holds, then statement (S1) of Lemma \ref{lemmix} is violated. So statement (S2) of  Lemma \ref{lemmix} must hold, which implies ``$[r_i,f(r_i, a_{-i})]$ FSD$_i $ $[\ell_i,f(\ell_i, b_{-i})]$" cannot hold, so (iii) of Definition \ref{nbap} fails to hold. This shows (ii) and (iii) of Definition \ref{nbap} cannot hold together, proving that $f$ satisfies NBA. \qed

The discussion following Definition \ref{nbap} shows that there can be a probabilistic mechanism that is not completely mixed, but which, still, satisfies NBA. The example below shows that the conclusion of Proposition \ref{prop1p} does not hold if agents do not have strict preference orderings over action-outcome pairs.

\subsection{An example}

\begin{ex} \label{ex1}  Consider an environment ${\cal E}$ with two agents $1,2,$ sets of actions $A_1=\{a^0,a^1\},$ $A_2=\{b^0,b^1\}$ and the set of outcomes is $Z=\{z^0,z^1\}.$ Consider the probabilistic mechanism $f:A_1\times A_2\rightarrow \Delta(Z)$ described as follows:
{\small{$$\mbox{ (I) }f((a^0,b^0),z^0)=1/3,f((a^0,b^0),z^1)=2/3,\mbox{ (II) }f((a^0,b^1),z^0)=1/2,f((a^0,b^1),z^1)=1/2,$$$$\mbox{ (III) }f((a^1,b^0),z^0)=1/3,f((a^0,b^0),z^1)=2/3,\mbox{ (IV) }f((a^1,b^1),z^0)=3/4,f((a^1,b^1),z^1)=1/4$$}}Note that $f$ gives the same probability distribution under $(a^0,b^0)$ and $(a^1,b^0),$ so $f(a^0,b^0)=f(a^1,b^0).$ 

Suppose the preference $\succsim_1$ of agent $1$ is: $(a^0,z^1)\sim_1 (a^1,z^0)\succ_1 (a^0,z^0)\sim_1 (a^1,z^1),$ so $\succsim_1$ is not a strict preference relation. Observe that 
$$\phi^{[a^0,f(a^0,b^0)]}_1(a^0,z^1)=2/3>\phi^{[a^1,f(a^1,b^0)]}_1(a^0,z^1)=1/3,$$
$$\phi^{[a^0,f(a^0,b^0)]}_1(a^1,z^0)=2/3>\phi^{[a^1,f(a^1,b^0)]}_1(a^1,z^0)=1/3,$$
$$\phi^{[a^0,f(a^0,b^0)]}_1(a^0,z^0)=\phi^{[a^1,f(a^1,b^0)]}_1(a^0,z^0)=1,$$
$$\phi^{[a^0,f(a^0,b^0)]}_1(a^1,z^1)=\phi^{[a^1,f(a^1,b^0)]}_1(a^1,z^1)=1.$$
This shows that $[a^0,f(a^0,b^0)]$ FSD$_1$ $[a^1,f(a^1,b^0)].$ So conditions (i)-(ii) of Definition \ref{nbap} holds with $i=1,$ $\ell_i=a^0,$ $r_i=a^1$ and $a_{-i}=b^0.$ Next observe that  
$$\phi^{[a^1,f(a^1,b^1)]}_1(a^0,z^1)=3/4>\phi^{[a^0,f(a^0,b^1)]}_1(a^0,z^1)=1/2,$$
$$\phi^{[a^1,f(a^1,b^1)]}_1(a^1,z^0)=3/4>\phi^{[a^0,f(a^0,b^1)]}_1(a^1,z^0)=1/2,$$
$$\phi^{[a^1,f(a^1,b^1)]}_1(a^0,z^0)=\phi^{[a^0,f(a^0,b^1)]}_1(a^0,z^0)=1,$$
$$\phi^{[a^1,f(a^1,b^1)]}_1(a^1,z^1)=\phi^{[a^0,f(a^0,b^1)]}_1(a^1,z^1)=1.$$
This shows that $[a^1,f(a^1,b^1)]$ FSD$_1$ $[a^0,f(a^0,b^1)].$ So condition (iii) of Definition \ref{nbap} also holds with $i=1,$ $\ell_i=a^0,$ $r_i=a^1$ and $b_{-i}=b^1.$ This shows BA holds for $f$, although $f$ is completely mixed. \end{ex}

\section{Some concluding remarks}

\label{sec5}

\hyperref[sec:sec5]{}

We have constructed a simple framework for analyzing aggregation problems with expressive preferences and have presented some initial results focussing on the phenomenon we call ``Brexit anomaly". Of course this opens up a number of further questions. Some immediate aspects worth looking into are, for example, what happens when agents might extend preferences to lotteries in various ways different from first order stochastic dominance; whether multi-stage mechanisms might help avoiding Brexit-anomaly etc. A more important aspect is, of course, analyzing a {\em{general}} problem for implementation when agents' preferences over the action-outcome pairs are private information, an example of which we provided in the introductory section.

\section*{Appendix}

{\bf Proof of Observation \ref{ob1}} First consider $1.$ To verify (i) of Definition \ref{sep} for $\succsim_1,$ first suppose 
$(x_1,(w,t)) \succ_1 (x_1,(\tilde{w},\tilde{t})).$ Since the distance between $1$'s report and its true waiting cost is $|x_1-\theta_1|$ for both of these action-outcome pairs, by properties (a)-(c) of $\succsim_1,$ we must have either [$u_1(w,t)>u_1(\tilde{w},\tilde{t})$] or  [$u_1(w,t)=u_1(\tilde{w},\tilde{t})$ and $t_1+t_2>\tilde{t}_1+\tilde{t}_2$]. In either case we have $(\tilde{x}_1,(w,t)) \succ_1 (\tilde{x}_1,(\tilde{w},\tilde{t}))$ for any $\tilde{x}_1.$ 

Next suppose 
$(x_1,(w,t)) \sim_1 (x_1,(\tilde{w},\tilde{t})).$ Then by property (d) of $\succsim_1,$ we must have $u_1(w,t)=u_1(\tilde{w},\tilde{t})$ and $t_1+t_2=\tilde{t}_1+\tilde{t}_2,$ implying $(\tilde{x}_1,(w,t)) \sim_1 (\tilde{x}_1,(\tilde{w},\tilde{t}))$ for any $\tilde{x}_1.$ This shows (i) of Definition \ref{sep} holds for $\succsim_1$.

To verify (ii) of Definition \ref{sep} for $\succsim_1,$ first suppose 
$(x_1,(w,t)) \succ_1 (\tilde{x}_1,(w,t)).$ These two action-outcome pairs have the same material payoff $u_1(w,t)$ for $1$ and the same revenue $t_1+t_2.$ So by property (c) of $\succsim_1,$ we must have $|x_1-\theta_1|>|\tilde{x}_1-\theta_1|.$ Applying properties (a)-(c) of $\succsim_1,$ we have $(x_1,(\tilde{w},\tilde{t})) \succ_1 (\tilde{x}_1,(\tilde{w},\tilde{t}))$ for any$(\tilde{w},\tilde{t}).$ 

Finally suppose 
$(x_1,(w,t)) \sim_1 (\tilde{x}_1,(w,t)).$ Then by properties (a)-(c) of $\succsim_1,$ we must have $|x_1-\theta_1|=|\tilde{x}_1-\theta_1|$ and property (d) implies 
$(x_1,(\tilde{w},\tilde{t})) \sim_1 (\tilde{x}_1,(\tilde{w},\tilde{t}))$ for any$(\tilde{w},\tilde{t}).$ This shows (ii) of Definition \ref{sep} holds for $\succsim_1,$ so $\succsim_1$ is separable. 

Now consider $2.$ For $2,$ the preference order between any two action-outcome pairs $(x_2,(w,t)),$ $(y_2,(\tilde{w},\tilde{t}))$ is solely determined by $(w,t),(\tilde{w},\tilde{t}).$ Specifically 
$$(x_2,(w,t)) \succsim_2 (y_2,(\tilde{w},\tilde{t}))\Leftrightarrow u_2(w,t)\geq u_2(\tilde{w},\tilde{t})$$Therefore $(x_2,(w,t)) \succsim_2 (x_2,(\tilde{w},\tilde{t}))$ implies $u_2(w,t)\geq u_2(\tilde{w},\tilde{t}),$ which in turn implies 
$(\tilde{x}_2,(w,t)) \succsim_2 (\tilde{x}_2,(\tilde{w},\tilde{t}))$ for any $\tilde{x}_2,$ so (i) of Definition \ref{sep} holds. 

Since $(x_2,(w,t)) \sim_1 (y_2,(\tilde{w},\tilde{t}))\Leftrightarrow $ $u_2(w,t)=u_2(\tilde{w},\tilde{t}),$ we have $(x_2,(w,t)) \sim_2 (\tilde{x}_2,(w,t))$ and $(x_2,(\tilde{w},\tilde{t})) \sim_2 (\tilde{x}_2,(\tilde{w},\tilde{t})),$ so (ii) of Definition \ref{sep} also holds, proving that $\succsim_1$ is separable. \qed

\section*{References}

\begin{description}

\item Austen-Smith, D., Banks, J.S. 1999. Positive Political Theory I. University of Michigan Press.

\item Borah, A. 2019. Voting expressively. Economic Inquiry. 57: 1617-1635.

\item Dutta, B., Sen, A. 2012. Nash implementation with partially honest individuals. Games and Economic Behavior. 74: 154-169.

\item Feddersen, T., Gailmard, S., Sandroni, A. 2009. Moral bias in large elections: Theory and experimental evidence. American Political Science Review. 103: 175-192.

\item Fishburn, P. 1970: Utility Theory for Decision Making. John Wiley.

\item Hillman, A. 2010. Expressive behavior in economics and politics. European Journal of Political Economy. 26: 403-418.

\item Kartik, N. 2009. Strategic communication with lying costs. Review of Economic Studies. 76: 1359-1395.

\item Kimya, M. 2017. Nash implementation and tie-breaking rules. Games and Economic Behavior. 102: 138-146.

\item Le Breton, M., Sen, A. 1999. Separable preferences, strategyproofness, and decomposability. Econometrica. 67: 605-628.

\item Lynskey, D. 2017. ‘I thought I’d put in a protest vote’: the people who regret voting leave. The Guardian, Nov 25, 2017. \url{https://www.theguardian.com/politics/2017/nov/25/protest-vote-regret-voting-leave-brexit} accessed April 4, 2021.

\item Mitra, M. 2001. Mechanism design in queueing problems. Economic Theory. 17: 277–305.

\item Moldovanu, B., Rosar, F. 2021. Brexit: A comparison of dynamic voting games with irreversible options. Games and Economic Behavior. 130: 85-108.

\item Suzumura K., Xu, Y. 2001. Characterizations of consequentialism and nonconsequentialism. Journal of Economic Theory. 101: 423-436.

\end{description}

\end{document}